# Beyond Technological Solutionism: Rethinking XR in Healthcare

A Critical Analysis of Care Coordination in Diabetes-CVD Management: A Systematic Design-Centered Approach

Md Haseen Akhtar†
Department of Design
Indian Institute of Technology
Kanpur, Uttar Pradesh, India
mohammad.haseenakhtar@fulbrightmail.org

Cecilia Landa-Avila
School of Design and Creative Arts
Loughborough University
Loughborough, Leicestershire, UK
I.C.Landa-Avila@lboro.ac.uk

Shital Desai
Scoail and Technological Systems Lab, School of Arts, Media, Performance and Design
York University, Toronto, Canada
desais@yorku.ca

Claire Warden
School of Design and Creative Arts
Loughborough University
Loughborough, Leicestershire, UK
C.Warden@lboro.ac.uk

Andrew Morris
School of Design and Creative Arts
Loughborough University
Loughborough, Leicestershire, UK
A.P.Morris@lboro.ac.uk

Mark Anderson
Department of Architecture, CED, University of California, Berkeley, USA
markand@berkeley.edu

Thomas Cochrane
Technology Enhanced Learning in Higher Education, CSHE, University of Melbourne, Australia
cochrane.t@unimelb.edu.au

Janakarajan Ramkumar
Department of Design
Indian Institute of Technology
Kanpur, Uttar Pradesh, India
jrkumar@iitk.ac.in

## ABSTRACT

The healthcare industry's enthusiastic adoption of Extended Reality (XR) technologies obscures a concerning reality: we were building increasingly sophisticated ways to perpetuate fundamentally broken healthcare systems. Through three deeply personal narratives - a rural patient cut off from care infrastructure, an urban professional navigating fragmented services, and a first-generation immigrant confronting cultural barriers - this provocation paper exposes how our obsession with technological innovation often worsens rather than resolves healthcare disparities. By applying the SEIPS 3.0 model to examine diabetes-CVD care coordination, we identify an "innovation paradox" where advanced technology creates new barriers to effective care. Our care interdependencies framework reveals that healthcare outcomes are shaped primarily by human relationships (50-60%), organizational coordination (25-30%), and sociocultural factors (15-20%), not technological sophistication. This research challenges the HCI community to confront its role in perpetuating healthcare inequities, demands a fundamental rethinking and proposes a new framework for healthcare innovation that prioritizes human relationships over technical capability, systemic change over feature sets, and actual care delivery over technological ambition. For healthcare providers, technology developers, and policymakers, our findings suggest that effective care coordination requires us to step back from our techno-solutionist mindset and engage with the messy reality of human-centered healthcare delivery.

## CCS CONCEPTS

## KEYWORDS

# 1 Introduction

When did we decide that more technology equals better healthcare? The UK healthcare system's staggering £4.2 billion burden from diabetes and cardiovascular disease [4] isn't just a financial crisis - it's evidence of a deeper failure in how we conceptualize healthcare innovation. While technologists and healthcare providers celebrate the rapid proliferation of digital health platforms [6], 4.3 million diabetes patients [3] find themselves trapped in what we call the "innovation paradox": the more sophisticated our solutions become, the more alienated patients feel from their own care.

Our research with Sarah (62), Katie (45), and Raven (51) exposes an uncomfortable truth: the healthcare technology sector has become complicit in perpetuating systemic inequities under the guise of innovation [7]. Consider these revelations from our study: Sarah's rural healthcare journey reveals how our obsession with "cutting-edge" solutions actively excludes communities that lack basic digital infrastructure; Katie's struggle with fragmented care shows how each new healthcare platform adds another layer of complexity to an already Byzantine system; and Raven's experience demonstrates how our technical "solutions" often embed cultural biases that create new barriers to care. The Systems Engineering Initiative for Patient Safety (SEIPS 3.0) model [15] was meant to put human experience at the center of healthcare design and describing the spatio-temporal distribution multiple care interactions. Instead, we have used it to justify increasingly complex technological interventions that ignored fundamental human needs. Our research reveals three critical failures in current healthcare innovation:

1. **The Accessibility Myth:** We design for an imaginary "average" patient while ignoring the diverse realities of healthcare access
2. **The Integration Fantasy:** We keep building new platforms while failing to address the fundamental disconnects in healthcare delivery
3. **The Innovation Trap:** We measure success by technological sophistication rather than human impact

This isn't just another critique of healthcare technology - it is a call to confront how our pursuit of innovation often becomes an excuse to avoid addressing systemic failures in healthcare delivery. Through our examination of three distinct patient journeys, we expose not just the limitations of current approaches, but the urgent need to fundamentally rethink what we mean by "innovation" in healthcare. Our work with three distinct patient profiles reveals a provocative truth: current technological solutions often reinforce rather than resolve healthcare disparities [7]. This raises uncomfortable but necessary questions about our approach to healthcare innovation:

1. Why do we continue to design healthcare technologies that ignore the rich complexity of human experience described in the SEIPS 3.0 model [15]?
2. How has our focus on technological capability overshadowed the need for meaningful care coordination [9]?
3. When did we decide that digital connection equals effective and safety care [14]?

While previous studies have applied SEIPS principles to healthcare system design [16], its application in conjunction with emerging technologies like XR remains understudied. This research gap is particularly significant in the context of chronic disease management, where the complex interplay of patient journeys, healthcare systems, and technological interventions requires systematic investigation [5].

# 2 Research Provocations

Instead of traditional research questions, we pose these provocative challenges to the HCI community:

***1. How have we failed to implement truly human-centered patient journey design across diverse populations?*** How have we failed to create genuinely inclusive patient journey designs that address the diverse realities of healthcare access? In what specific ways have our design approaches, methodologies, and assumptions systematically overlooked or marginalized the experiences of patients across different geographic settings, professional contexts, and cultural backgrounds? What fundamental disconnects exist between our human-centered design rhetoric and the lived experiences of patient’s persona like Sarah, Katie, and Raven?

***2. What critical blind spots persist in our understanding of diverse stakeholder perspectives that systematically undermine care coordination efforts?*** Which voices and experiences remain marginalized or entirely absent from our design processes, and how does this perpetuate fragmentation in healthcare delivery? How might our limited engagement with the full spectrum of healthcare stakeholders - from rural patients to cultural mediators to time-constrained clinicians - prevent us from identifying the true barriers to effective care coordination?

***3. Why do we assume that immersive technology will solve problems that fundamentally stem from human system failures?*** Why do we persistently invest in immersive technology as a solution to healthcare challenges that fundamentally stem from human system failures? What underlying assumptions drive our faith that technological sophistication will resolve issues that are primarily organizational, relational, or structural in nature? How might our focus on creating novel XR experiences distract us from addressing the deeper systemic problems in healthcare coordination that technology alone cannot solve?

## 3 Methodology: Reflections and Learnings

Our methodological journey exposed three fundamental failures in how we approach healthcare research. First, our initial attempt to use traditional research methods nearly obscured the very insights we sought to uncover. The academic pressure to produce 'statistically significant' results almost led us to sacrifice the deep understanding that only comes from sustained engagement with individual experiences. Our journey with Sarah, Katie, and Raven revealed how conventional research approaches often perpetuate the very problems they claim to solve.

Our journey through this research challenged conventional wisdom about studying healthcare technology integration. The decision to focus on just three cases - Sarah (62), Katie (45), and Raven (51) - initially faced skepticism from reviewers who questioned statistical significance. However, this focused approach revealed something profound: the depth of understanding gained from sustained engagement with fewer participants far outweighed the superficial insights from larger sample sizes [21, 22]. Each case represents a unique contextual variation (rural, urban-professional, cultural-linguistic) while sharing the common thread of diabetes-CVD comorbidity management. All three participants (Sarah, 62; Katie, 45; and Raven, 51) live with long-term comorbidities of diabetes and cardiovascular disease, interacting with different healthcare sub-services based on their unique contexts. Sarah primarily engages with rural primary care and emergency services from her isolated location, facing significant geographic barriers and digital infrastructure limitations. Katie navigates between corporate healthcare services and multiple specialist providers across three different hospitals, balancing her treatment with a high-pressure financial sector career where health issues are stigmatized. Raven's healthcare journey involves primary care and hospital services complicated by cultural-linguistic barriers as a first-generation immigrant, requiring family members for translation and incorporating traditional healing practices alongside western medicine.

**What we got right:** Using journey mapping as a visual dialogue tool transformed our interviews from simple data collection into collaborative meaning-making sessions [17]; Conducting multiple sessions over three months allowed trust to build, revealing deeper insights about healthcare coordination challenges; and, Embracing the complexity of each case rather than trying to standardize the research process [23].

**What we got wrong:** Initially attempting to fit diverse healthcare experiences into predetermined categories; Underestimating the emotional labor involved in discussing healthcare challenges; and, assuming that technology literacy was the primary barrier to XR adoption [10].

## 4 Discussion

The healthcare technology sector's obsession with data collection has become a dangerous distraction from what matters in patient care. Our analysis exposes a fundamental delusion: the belief that more data somehow equals better healthcare. This "data-first" mindset has created what we term the "measurement paradox" - the more we measure, the less we understand about actual patient experiences [28]. Our weighted dependencies framework emerged not from a desire to create another classification system, but as damning evidence of how thoroughly we've misunderstood healthcare delivery. The framework exposes three levels of systemic failure: While the technology sector chases the next breakthrough in digital health platforms, our research reveals that 50-60% of healthcare outcomes depend on something far more basic: human relationships (The Human Element We Keep Ignoring). The success of healthcare interventions, particularly in complex cases involving multiple conditions, hinges not on technological sophistication but on the quality of human connections [28]. Yet we continue to invest in solutions that often erode rather than enhance these crucial relationships. The promise of "seamless integration" across healthcare organizations has become a hollow marketing slogan. Our analysis shows that 25-30% of healthcare delivery effectiveness depends on inter-organizational coordination (The Integration Myth) [29], yet each new technological solution seems to add another layer of complexity to an already fragmented system. We've created what we call "digital silos" - sophisticated technical solutions that make it harder for healthcare organizations to work together effectively. Perhaps most damning is our systematic neglect of sociocultural factors in healthcare delivery. While these "tertiary" dependencies typically account for 15-20% of impact (The Cultural Blind Spot) [17], our research reveals how they can become decisive in specific contexts. Rural communities and culturally diverse populations don't need more sophisticated apps - they need healthcare systems that acknowledge their fundamental human realities. This isn't just about improving existing systems; it's about confronting how our techno-centric approach to healthcare innovation has become part of the problem. When a rural patient like Sarah can't access basic healthcare services, adding another layer of digital complexity isn't just unhelpful - it's actively harmful. When Raven struggles with cultural barriers to care, offering him another app to navigate becomes an exercise in technological theater rather than meaningful healthcare delivery. The weighted dependencies framework (Figure 1) doesn't just describe healthcare delivery - it indicts our current approach to healthcare innovation. It exposes how we have systematically invested in the wrong solutions, chasing technological sophistication while ignoring the human foundations of effective healthcare delivery [29].

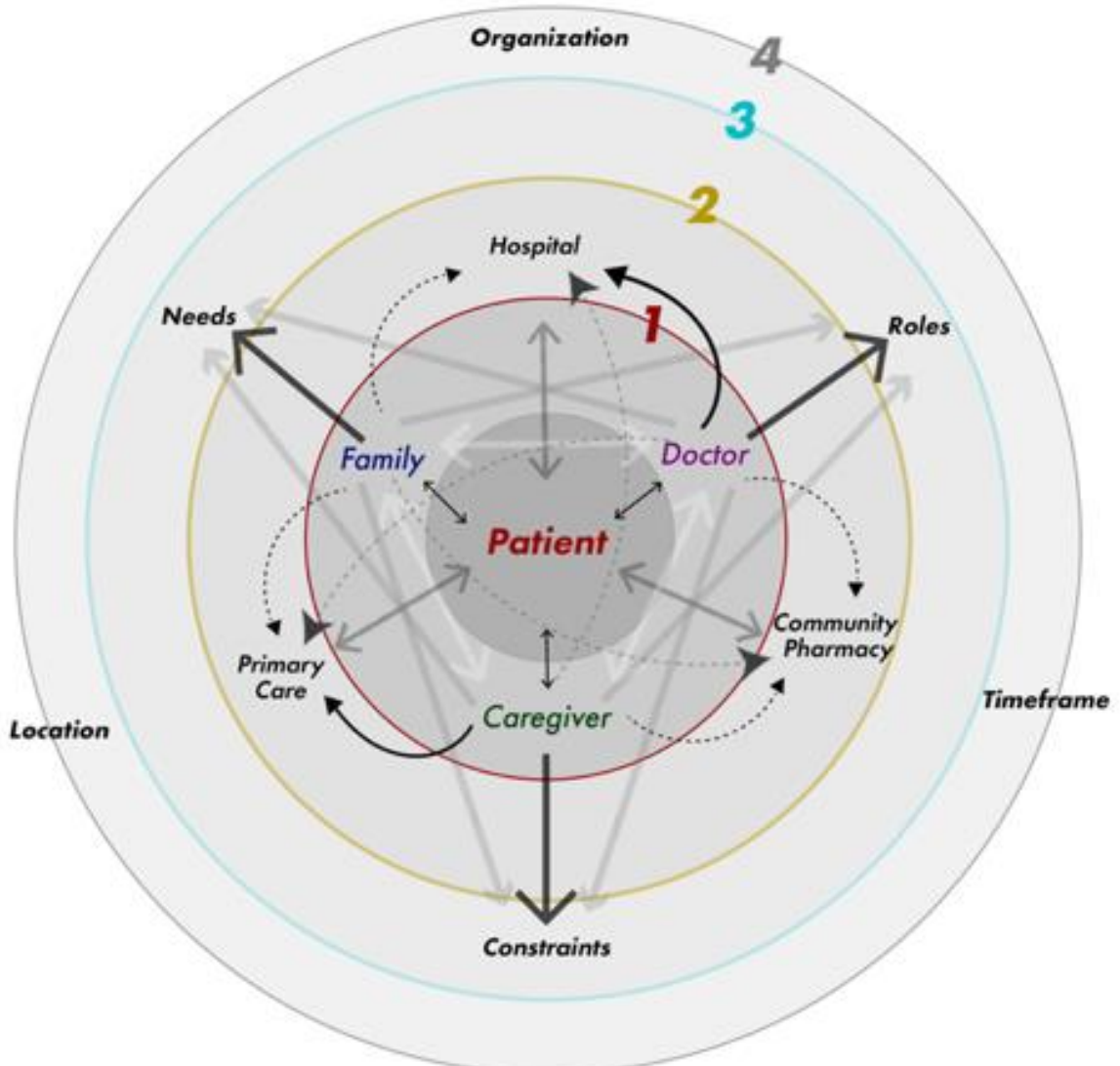

**Figure 1: Care Interdependencies Framework; Primary Dependencies (within 1 - marked by red) – Direct Patient Provider Relationships, Care Delivery Effectiveness (50-60%); Secondary Dependencies (within 1 and 2 - marked by yellow) – Healthcare Organization Coordination, Resource Sharing (25-30%); and Tertiary Dependencies (within 2 and 3 – marked by blue) - Cultural, Geographic and Social Influences (15-20%)**

## 5 Insights: Truths about HCI in Healthcare

We found that the concept of "universal design" in healthcare technology is fundamentally questionable. Sarah's rural isolation, Katie's professional demands, and Raven's cultural context weren't just variables to control for - they were fundamental aspects of their healthcare experience that demanded distinctly different technological approaches [2]. The weighted interdependencies framework we developed reveals how these contexts fundamentally shape technology effectiveness. While the healthcare technology sector promotes "seamless integration," our research exposes this as largely aspirational [28]. The weighted interdependencies framework reveals three distinct levels of dependencies:

1. **Primary Dependencies** (50-60% impact): Direct patient-provider relationships
2. **Secondary Dependencies** (25-30% impact): Inter-organizational coordination
3. **Tertiary Dependencies** (15-20% impact): Sociocultural and environmental factors

Perhaps most provocatively, we found that more sophisticated technology doesn't necessarily lead to better care coordination [29]. In fact, there were instances where high-tech solutions increased the coordination burden by adding layers of complexity to already strained healthcare workflows. This research isn't just about understanding healthcare coordination - it's about challenging the HCI community to fundamentally rethink its approach to healthcare innovation. We propose three provocative principles for future work:

***1. Embrace Complexity:*** Stop trying to simplify healthcare coordination into neat technological solutions. Our weighted interdependencies framework shows that successful healthcare innovation requires engaging with the messy reality of human-centered care [15]. Healthcare journeys, particularly for patients with comorbidities like diabetes and CVD, resist linear, predictable pathways that technologists often try to impose. Our research with Sarah, Katie, and Raven reveals how patient journeys are inherently shaped by overlapping contexts - geographic limitations, professional demands, cultural practices - that cannot be reduced to standardized workflows or algorithmic solutions. This principle challenges us to design systems that accommodate ambiguity, adaptation, and the inherent unpredictability of human health experiences. Rather than creating 'cleaner' interfaces that hide complexity, we should develop technologies that make complexity navigable, visible, and manageable for both patients and providers. This means embracing design approaches that allow for flexibility, supporting healthcare decisions that often occur in contexts of uncertainty, and developing solutions that enhance human judgment rather than attempting to replace it with technological determinism. By acknowledging complexity as a feature rather than a bug in healthcare coordination, we open possibilities for more resilient, adaptive systems that better reflect the actual experiences of patients and providers."

***2. Challenge Technological Solutionism:*** Not every healthcare coordination challenge needs a high-tech solution. Sometimes, the most effective interventions involve redesigning existing processes rather than adding new technological layers [26]. This principle directly confronts our 'data-first' mindset and the resulting 'measurement paradox' we identified earlier—where more data collection often leads to less understanding of patient experiences [28]. As healthcare systems increasingly push toward data-driven approaches, we must recognize that data quantity doesn't automatically translate to care quality. Our research shows that excessive measurement without meaningful integration often fragments rather than enhances coordination. Rather than continuously adding more sensing, tracking, and monitoring technologies that burden both patients and providers, we should focus on improving how existing information flows between human actors in the healthcare ecosystem. True innovation might sometimes mean fewer technological layers and more thoughtful process redesign."

***3. Prioritize Integration over Innovation:*** The HCI community must shift its focus from creating novel technologies to understanding and supporting the complex web of human relationships that make healthcare work [25]. Our care dependencies framework reveals that 50-60% of healthcare outcomes depend on these human relationships, yet our technological innovations frequently disrupt rather than enhance these connections. True progress in healthcare delivery requires us to redirect our creative energy toward integrating existing systems and strengthening human connections rather than developing ever-newer technologies that increase complexity. This means designing solutions that enhance communication between providers, bridge organizational boundaries, and support family care networks - prioritizing the continuity of relationships over technological novelty. By focusing on integration, we acknowledge that healthcare is fundamentally a human system supported by technology, not a technological system that happens to involve humans. This principle challenges us to measure success not by technological sophistication or feature sets, but by how effectively our solutions strengthen the human relationships that form the foundation of effective care." Failed papers teach us:

- Rejected works often focused too heavily on technological capability rather than healthcare impact
- Successful papers engaged deeply with the human aspects of healthcare coordination
- The most impactful work challenged existing assumptions about healthcare technology

The future of HCI in healthcare lies not in creating more sophisticated technologies, but in better understanding and supporting the human systems that make healthcare work. This requires a fundamental shift in how we approach healthcare innovation, moving beyond technological solutionism to embrace the complex reality of human-centered care coordination.